\documentclass[a4paper,conference]{IEEEtran}
\IEEEoverridecommandlockouts

\IEEEsettopmargin{t}{30mm}
\IEEEquantizetextheight{c}
\IEEEsettextwidth{14mm}{14mm}
\IEEEsetsidemargin{c}{0mm}
\usepackage{url}
\usepackage{ifthen}
\usepackage[cmex10]{amsmath} 
\usepackage[T1]{fontenc}
\usepackage[utf8]{inputenc}

\usepackage{amssymb,amsfonts,mathtools,physics}

\usepackage[dvipsnames]{xcolor}
\usepackage{graphicx}

\usepackage{tikz,tikzscale}
\graphicspath{{figures/}}

\makeatletter
\newcommand{\AddInputPath}[1]{%
    \providecommand*{\input@path}{}
    \g@addto@macro{\input@path}{#1}
}
\makeatother
\AddInputPath{{figures/}}

\usepackage{hyperref}
\usepackage[capitalize]{cleveref}

\newcommand{\field}{\mathbb{F}}
\newcommand{\naturals}{\mathbb{N}}
\newcommand{\eps}{\varepsilon}

\begin{document}

\title{Reed-Muller Identification
\thanks{We acknowledge support from the German Federal Ministry of Education and Research (BMBF) to C. Deppe and R. Ferrara under Grant 16KIS1005.
Contact information: \{roberto.ferrara,christian.deppe\}@tum.de}
}

\author{\IEEEauthorblockN{Mattia Spandri, Roberto Ferrara, Christian Deppe}
\IEEEauthorblockA{\textit{Institute for Communication Engineering},
\textit{Technical University of Munich}, Munich, Germany \\
}
}

\maketitle

\begin{abstract}
Ahlswede and Dueck identification has the potential of exponentially reducing traffic or exponentially increasing rates in applications where a full decoding of the message is not necessary and, instead, a simple verification of the message of interest suffices.
However, the proposed constructions can suffer from exponential increase in the computational load at the sender and receiver, rendering these advantages unusable.
This has been shown in particular to be the case for a construction achieving identification capacity based on concatenated Reed-Solomon codes.
Here, we consider the natural generalization of identification based on Reed-Muller codes and we show that they achieve identification capacity and allow to achieve the exponentially large rates mentioned above without the computational penalty increasing too much the latency with respect to transmission.
\end{abstract}


\section{Introduction}
    Ahlswede and Dueck's identification is a different communication paradigm from Shannon's transmission that promises an exponential larger capacity, or equivalently an exponential reduction in channel uses, at the trade-off of only allowing an hypothesis test at the receiver instead of a full decoding~\cite{AD89}.
    Identification capacity on a noisy channel can be achieved by concatenating a capacity-achieving identification code for the noiseless channel with a capacity achieving transmission code for the noisy channel~\cite{AD89}.
    In other words, it is enough to correct the channel first and then apply some pre and post processing.
    
    As common for capacity results, the achievability proof ignores the complexity of constructing the code and complexity of the encoder and decoder.
    In particular, since identification promises an exponential increase in the rates, even simply reading the chosen identity (sometimes still called message to make the parallel with transmission) will incur some penalty.
    In previous works~\cite{DDF20,FTBDLMV21}, we analyzed the time spent encoding and the noiseless-channel error probability for a capacity-achieving noiseless identification codes based on concatenated Reed-Solomon codes~\cite{VW93explicit}.
    The result from those works was that, with todays transmission speeds, it is generally faster to simply send the unique string defining the identity than spend the time encoding for identification.
    In order to make noiseless identification competitive in terms of latency, the use of Zech tables was necessary to speed up the computation over finite fields, however this option was limited to codes of small size, leaving the open question of finding similarly fast identification-codes at larger sizes.
    The codes from~\cite{VW93explicit} are only one of possible identification capacity-achieving constructions, which can generally be obtained via block codes satisfying the Gilbert-Varshamov bound~\cite[Section~III.B]{AZ95newdirections}.
    Other such constructions are the algebraic codes of~\cite{KTV84,EZ87} as pointed out in~\cite{VW93explicit}, a construction based on hash functions~\cite{KY99hashID}, and the recent construction of~\cite{GKSS21}.
    
    In this work, we naturally generalize to identification codes base on Reed-Muller codes in order to increase the size of the identities without increasing the size of the finite fields we work on.
    We find that, although requiring small field sizes also limits how low we can make the error probability, we can circumvent this using multiple encoding~\cite{FTBDLMV21} and efficiently reduce the error without much impact on the other parameters.
    
    The paper is structured as follows.
    In \cref{sec:ID,sec:RM}, we quickly review identification and Reed-Muller codes.
    In \cref{sec:capacity}, we show that they alone can achieve identification capacity without concatenation.
    In \cref{sec:performance}, we discuss the implementation and show how it allows to achieve large exponential increase in rates without much latency and false-accept penalty compared to transmission.
    In the appendices, we describe in detail how we measured the time cost of operations in an attempt to predict the performance of the code.

\section{Identification}
\label{sec:ID}
    We use the notation $[n]=\{0,...,n-1\}$ for any natural number $n$ and the notation $PW = \sum_{x} P(x) W(\cdot|x)$ for a probability distribution $P$ and a channel $W$.
    An $(n,I,\eps)$ identification code for $W^n$ is a tuple $\{E_i, V_i\}_{i\in[I]}$ of probability distributions and verifier sets (like stochastic codes for transmission) such that $e_{ij} = \qty|E_iW^n(V_j) - \delta_{ij}|\leq \eps$, where $\delta_{ij}$ is the Kronecker delta (notice that $e_{ii}$ are the usual errors in transmission).
    No disjointness or limit on the intersection is imposed on the verifier sets.
    The rate is defined as $\frac{1}{n}\log \log I $ rather than $\frac{1}{n} \log I$, and the capacity is then the supremum of achievable rates as usual.
    As mentioned already, we can focus only on coding for the noiseless channel, like in~\cite{FTBDLMV21}, in which case it is enough to construct the appropriate verifier sets $V_i$ and let $E_i$ be the uniform distributions on these sets~\cite{AD89}.
    One way to do this, is to construct the verifier sets sets using functions $f_i:[R] \to [T]$ for each identity $i$, such that $[R]\times [T] =[RT]$ can be mapped one-to-one to the inputs of the noiseless channel. 
    These sets are then none other than the relation sets $V_i = \{(r,f_i(r))\}_{r\in[R]} \subset [R] \times [T]$ defined by $f_i$.
    We call $r$ the \emph{randomness} and $f_i(r)$ the \emph{tag}.
    By construction, we can then think of the encoder as choosing a random challenge in the form of a randomness-tag pair $(r,f_i(r))$ and sending it through the channel, so that the receiver wanting to verify identity $j$ will recompute the tag $f_j(r)$ and conclude that $i=j$ if the recomputed tag is equal to the received tag $f_j(r) \stackrel{?}{=} f_i(r)$~\cite{AD89feedback,MK06,DDF20,FTBDLMV21}.
    With such a scheme $e_{ii}$ will always be $0$, while $e_{ij}$ is bounded by the fraction of collisions (outputs that coincide) of $f_i$ and $f_j$.
    To limit $e_{ij}$, the number of collisions needs to be limited, which makes the set of such identification codes in one to one correspondence with error-correction block codes: each codeword (a string of symbols) defines a function from symbol positions to symbol values and the distance of the code gives a bound on the false-accept error probability~\cite{DDF20}.
    For example, using Reed-Solomon codes, the functions corresponding to the codewords are none other than the polynomials used to generate the codewords.

\section{Reed-Muller Codes} 
\label{sec:RM}
    For our purpose, it will make sense to consider $q$-ary rather than just binary Reed-Muller code~\cite{KLP68,DGM70,MCJ73}.
    Let $k,m\in\naturals$ and $q>k$ a prime power.
    Because of our application to identification, we define the $\mathrm{RM}_{q}(k,m)$ Reed-Muller code as the collection of multivariate polynomials with $m$ variables and degree at most $k$ over $\field_q$.
    For this, we introduce some notation first.
    For any vector of exponents $z\in[k]^m$ and any vector of variables $r\in\field_q^m$, we define the degree $|z| \coloneqq \sum_{j=1}^{m} z_j$ and the monomial $r^z \coloneqq \prod_{j=1}^{m} r_j^{z_j}$.
    The Reed-Muller code is then defined as 
    \begin{align}
    \mathrm{RM}_{q}(k,m) = 
    \qty{
    \begin{aligned}
    &p_w:\field_q ^m \to \field_q 
    \\
    &p_w(r) = \sum\nolimits_{{z\in[k]^m: |z| \leq k}} w_z r^z 
    \end{aligned}},
    \end{align}
    where $w = \qty{w_z}_{{z\in[k]^m: |z|\leq k}}$ are the $\binom{k+m}{m}$ coefficients in $\field_q$.\\
    In case of a Reed-Muller error-correction code then every polynomial constructs a codeword by concatenating polynomial evaluations at different input points.
    The maximum blocklength is of course the number of possible inputs, which results in a block code with parameters
    \begin{align}
    \left[q^{m},\binom{k+m}{m}, (q-k)q^{m-1}\right]_{q}.
    \label{eq:RM-blockcode}
    \end{align}
    In identification, a functional encoding is preferred so that a single letter of the codewords can be computed without computing the whole codeword~\cite{DDF20,FTBDLMV21}.
    For the Reed-Muller code, this is the polynomial encoding. 
    The size of the Reed-Muller identification code is the number of distinct polynomials, given in bits (all logarithms are in base two) by
    \begin{equation}
        \log I  = \binom{k+m}{m} {\log q}.
        \label{eq:I}
    \end{equation}
    However, only a transmission of 
    \begin{equation}
        \log C = (m+1) \log q
        \label{eq:C}
    \end{equation}
    bits is needed, since only the challenge, composed of 
    \begin{align}
        \log R  &= m \log q 
        &
        \text{and}&
        &
        \log T = \log q 
        \label{eq:R}
    \end{align}
    bits of randomness and tag, is sent through the channel.
    Thus for a single Reed-Muller code, 
    the increase from the transmission rate $r_\mathrm{T}$ to the identification rate $r_\mathrm{ID}$ is
    \begin{align}
    \frac{r_\mathrm{ID}}{r_\mathrm{T}} = \frac{\log I}{\log C}  = \frac{\binom{k+m}{m}}{m+1}.
    \end{align}
    compared to $\frac{r_\mathrm{ID}}{r_\mathrm{T}} =  \frac{k}{2}$ of a Reed-Solomon code~\cite{MK06,DDF20}. 
    If multiple $n$ challenges are sent, this reduces the error but also reduces the rate increase to ${\binom{k+m}{m}}/{n/(m+1)}$. 
    The errors $e_{ij}$ are upper bounded by the fractional distance
    \begin{align}
        E = 1 - \frac{(q-k)q^{m-1}}{R} = \frac{k}{q}.
        \label{eq:E}
    \end{align}
    This is independent of the number of variables $m$ and less than one because $k<q$. 
    The error decreases as $E^n = (\frac{k}{q})^n$ with the number of challenges, because all challenges need to be verified simultaneously.

\section{Capacity}
\label{sec:capacity}

    In order to achieve identification capacity, a noiseless identification codes need to satisfy three simple conditions~\cite{VW93explicit}%
    \footnote{In~\cite{VW93explicit}, these conditions are called ``optimal'' for identification in the sense of achieving capacity, but not in the sense of optimal at finite blocklengths.}: 
    \begin{enumerate}
        \item 
        Randomness: asymptotically all the transmission rate is used for randomness:
        \begin{align}
            \label{id:randomness}
            \frac{\log T}{\log R} &\to 0
            &
            \Leftrightarrow&
            &
            \frac{\log R}{\log RT} &\to 1
        \end{align}
        where $C=RT$ is the size of the challenge;
        \item 
        Rate: asymptotically the identification rate must equal the randomness/transmission rate:
        \begin{equation}
            \label{id:rate}
            \frac{\log\log I}{\log R} \to 1;
        \end{equation}
        \item 
        Error: asymptotically the error must go to zero
        \begin{equation}
            \label{id:error}
            E = 1 - \frac{1}{R} \max_{i\neq j} d(T_i,T_j) \to 0.
        \end{equation}
    \end{enumerate}
    Since Reed-Muller codes contain Reed-Solomon codes as a special case, they are also able to achieve identification capacity using concatenation of multiple codes.
    The question is whether capacity can be achieved without concatenation, which is not possible with Reed-Solomon codes~\cite{VW93explicit}.
    Below we give a sequence of parameters of the Reed-Muller codes that satisfies these three properties, showing that they can achieve the identification capacity of the noiseless channels.
    
\subsection{Capacity-achieving sequence}    
    We begin with the randomness, \cref{id:randomness}, which requires
    \[\frac{\log T}{\log R} = \frac{1}{m} \to 0 
    \qquad \Rightarrow \qquad
    m\to\infty \]
    independent of the number of challenges. 
    From the error requirement, \cref{id:error}, we need to satisfy
    \begin{align}
        E &= \frac{k}{q} \to 0
        &
        &\Rightarrow
        &
        q &\to \infty.
        \label{eq:nlog-q}
    \end{align}
    Now we can compute the rate under these conditions and then choose $q,k,m$ appropriately. 
    For the rate, \cref{id:rate}, we get
    \begin{align*}
        \frac{\log\log I}{\log R} 
        &
        = \frac{\log\log q + \log \binom{k+m}{m}}{m \log q} 
        \to \frac{\log \binom{k+m}{m}}{m \log q} 
        \intertext{where we used that $\log\log q/ \log q \to 0$ by \cref{eq:nlog-q}.
        Now we can use the upper and lower bounds on the binomial $\qty(\frac{ a}{b})^b\leq\binom{a}{b}\leq \qty(\frac{\mathrm{e} a}{b})^b$ with $\mathrm{e}$ Euler's number. 
        We thus bound}
        \frac{\log \binom{k+m}{m}}{m \log q} 
        &
        \in  \qty[\frac{\log \frac{k+m}{m}}{\log q}, \frac{\log \frac{\mathrm{e}(k+m)}{m}}{\log q} ]
        \to \frac{\log \frac{k+m}{m}}{\log q}
    \end{align*}
    since $\log \mathrm{e} / \log q \to 0$.
    We can now choose
    \begin{align*}
        q &= 2^{t^2} 
        &
        k &= 2^{t^2-t}
        &
        m &= 2^t
    \end{align*}
    which implies $\frac{k+m}{m} \to \frac{k}{m}$ and gives 
    \begin{align*}
        \frac{\log\log I}{\log R} 
        &\to \frac{\log \frac{k}{m}}{\log q}
        = \frac{t^2 - 2t}{t^2} \to 1,
    \end{align*}
    proving capacity.

    Notice however, that in order to achieve identification capacity $q$ must grow to infinity, which directly clashes with our goal of keeping $q$ bounded.
    Even if we allow multiple challenges $n$, bounding $q$ implies bounding $k$ and since $m$ still needs to go to infinity, the tighter upper bound on the binomial becomes $\binom{k+m}{m}=\binom{k+m}{k} \leq (\mathrm{e} \frac{k+m}{k})^k$ which gives
    \begin{align*}
        \frac{\log\log I}{\log R} 
        &
        \leq  \frac{\log\log q + k \log \mathrm{e}\frac{k+m}{k}}{nm \log q} 
        \in O\qty(\frac{\log m}{mn}) 
    \end{align*}
    Thus with this constraint we cannot achieve any positive double-exponential rate, but only positive rates at scaling of the form $\frac{\log I}{N^\alpha}$ where $N$ is the blocklength and is a constant $\alpha < k$ which depends on $k$ and the scaling of $n$.

\section{Performance}
\label{sec:performance}
    \begin{figure}
        \centering
        \includegraphics[width = \columnwidth] {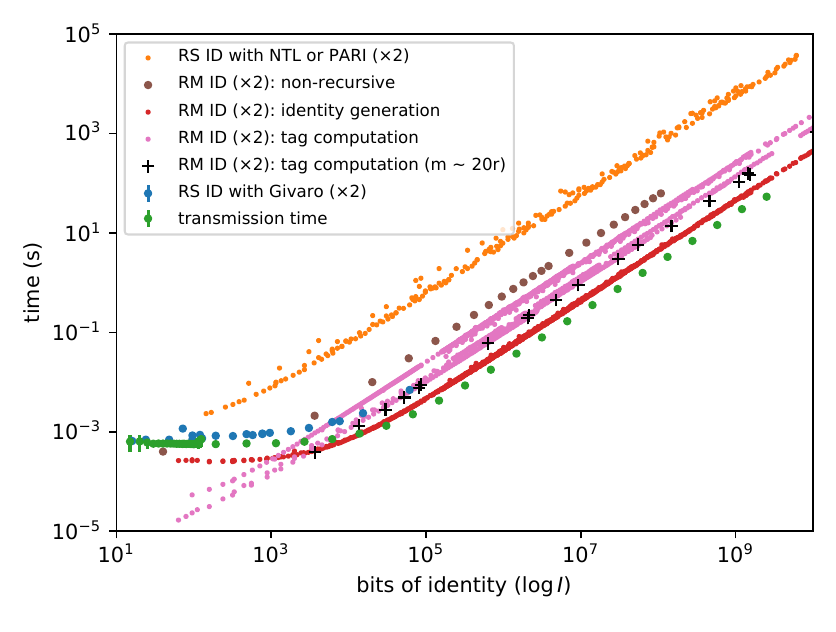}
        \caption{Time cost of identity generation (red) and non-optimized (brown) and optimized encoding (pink and black) for Reed-Muller identification codes compared to the data from~\cite{FTBDLMV21}: the cost of generation and encoding for concatenated Reed-Solomon identification codes (blue and orange) and the cost of direct transmission with an experimental setup (green).}
        \label{fig:RM-cost}
    \end{figure}
    
    \begin{figure}
        \centering
        {\includegraphics[width = \columnwidth, trim=0 25 0 25, clip] {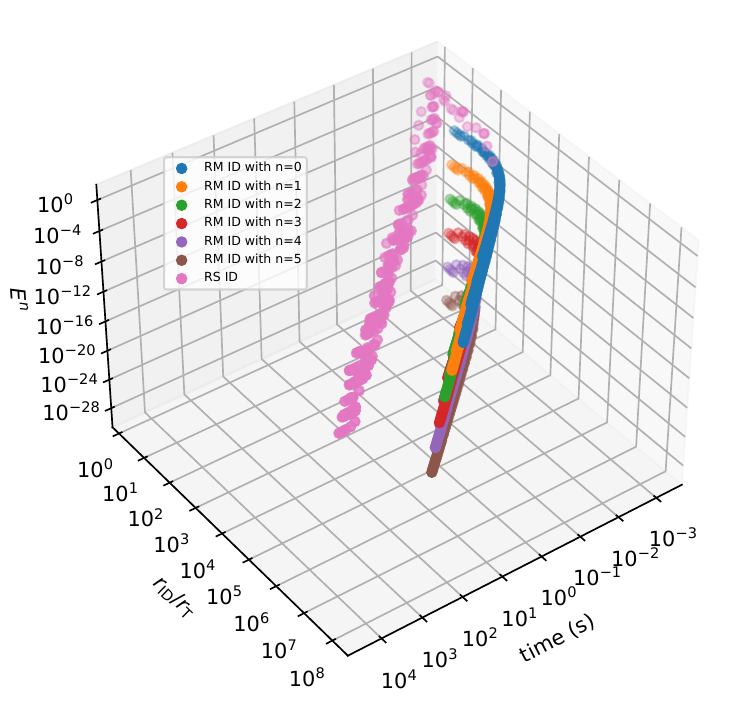}}
        \caption{Trade-off between time (generation, computation of $n$ challenges and transmission) the error and the size of the codes (shown for $n=1,...,6$) compared to the concatenated Reed-Solomon (pink points). The increased number of challenges successfully reduces the error without meaningfully impacting the computation time.} 
        \label{fig:RM-3d}
    \end{figure}
    
    Even though Reed-Muller codes achieve identification capacity only with large field sizes, they were still successful in our goal of implementing large identification codes with end-to-end time comparable with direct transmission and arbitrarily small error, as shown in figures \cref{fig:RM-cost,fig:RM-3d}.
    In order to achieve this performance, a combination of field size, computation optimization, and multiple challenges was used.

\subsection{Field size}
    As identified in ~\cite{FTBDLMV21}, the largest contribution to the computation time was the actual time of addition and multiplication operations in the Sagemath implementation. 
    Limiting the field size to $q<2^{16}$, where Zech tables of element logarithms are used, was the first step in achieving faster computation. 
    As shown in \cref{sec:GF_timing}, this led us to addition and multiplication times essentially equal and constant across any field size $q<2^{16}$.
    Thus, choosing the largest field within the constraint allows to increase the size in \cref{eq:I} and lower the error in \cref{eq:E} within our constraint.
    However, the bound $q<2^{16}$ also puts a lower bound on the error with a single challenge and thus multiple $n$ challenges need to be used to reduce the error exponentially.

\subsection{Computational optimization}
    The most efficient way of computing a polynomial is clearly to have it reduced into product of irreducible polynomials. 
    However, the cost of the reduction contributes to the identification encoding. 
    With a single variable, a fully reduced polynomial of degree $k$ 
    is computed in $2k-1$ operations.
    However, $2k$ operations are also achieved by computing the non reduced polynomial recursively. 
    From a programming point of view, recursion might introduce noticeable overhead and memory increase.
    Still, this means that we can optimize the number of operations without reduction.
    Recursion over degrees turned out to be too expensive and thus we use recursion only over variables as 
    \[p_w(r) = \sum\nolimits_{k'=[k]} r_1^{k'} \cdot p_{w_{k'}}(r_2...r_m),\]
    where $w_{k'}$ denote a partition of coefficients for polynomials of degree $k-k'$ and $m-1$ variables.
    This is none other than the Plotkin construction.
    Even then, the recursion turned out to be expensive as explained later below.
    
    The computation time (times two, since the tag must be computed at the sender and at the receiver) is plotted in \cref{fig:RM-cost} in pink and black. 
    For comparison, the brown points are the computation times without recursion.
    The improvement is larger than the caching optimization available for finite field computation (mentioned in \cref{sec:GF_timing}).
    The pink points actually form a band rather than a line, indicating that that there is room for optimization even in the choice of parameters $k$, $m$ ($q$ was already optimized as the largest $q<2^{16}$).
    Here is where we can see that the recursion still constitutes an expensive contribution for large $m$; the black point are a heuristic selection of parameters satisfying $k/m \in [10,50]$ indicating that the fastest computation happens for $k \gg m$.
    \Cref{sec:analytic} describes a failed attempt to analytically predict this behaviour and extract the optimal parameters.
    
    Finally, the red points represent the time spend in randomly generating the identities $w$.
    We have timed the generation and the encoding separately since this contribution might not be relevant depending on the application. 
    Since \cref{fig:RM-cost} is in log scale, the generation time is only a minor contribution.

\subsection{Multiple challenges}
    \Cref{fig:RM-cost} only shows size and computational time and thus does not show that the error of the Reed-Muller code increases with $k$ and thus the size (\cref{eq:E}), as opposed to the Reed-Solomon code where it decreases.
    Multiple challenges can be used to reduce the error~\cite{FTBDLMV21} at the cost of increasing computation time and transmission size.
    The decrease is exponential and thus only a small number of challenges is needed.
    The trade-off is displayed in \cref{fig:RM-3d} where the Reed-Muller code with a few challenges achieves points toward large size, small computation time and small error, more efficiently than the Reed-Solomon code.

\section{Conclusion}
    We have shown that it is possible to implement identification with latency comparable to current transmission speeds and arbitrarily small error.
    Better codes might even be able to be strictly faster than transmission in end-to-end identification.
    In particular, Polar codes are a potential candidate as they are characterized, among other things, by fast encoding times.
    Future work will also focus on verifying the advantage of identification in specific applications.
    Overall, our work shows that identification could potentially be an important technology in reducing traffic, load, latency in applications where the amount of data eventually grows faster than the capacity of the infrastructure.

\appendices

\section{Field additions and multiplications}
\label{sec:GF_timing}
    \begin{figure}
        \centering
        \includegraphics[width = \columnwidth] {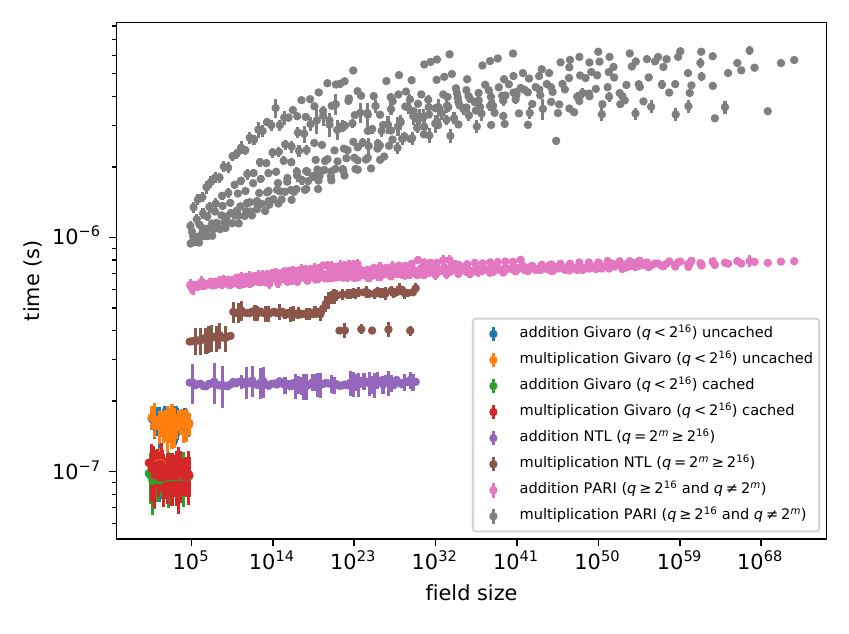}
        \caption{Above: Time cost of addition and multiplication for all field sizes, depending on the prime power either Givaro, NTL or PARI implementations are used by Sage. 
        With Givaro, there is an additional option to cache elements for  faster computation.
        }
        \label{fig:GF_timing}
    \end{figure}

    We measured the time spent performing additions and multiplications at various field sizes $q$.
    The results are shown in \cref{fig:GF_timing}.
    As expected, for $q\geq 2^{16}$ operation time increases considerably and multiplication time is noticeably larger than addition time.
    Multiplication and addition time is essentially the same for $q<2^{16}$ and, maybe unexpectedly, is independent of $q$.
    Such result suggests that the optimal choice is to choose the largest field size below $2^{16}$ in order to reduce the error.
    
    Finally, for $q<2^{16}$ there is an option to cache field elements, which seems to improve operation times uniformly by a factor $\sim 0.6$.
    As seen in \cref{sec:performance,sec:analytic}, other contributions influence the computation time more than the cache, making this factor not particularly relevant at the moment.
    We also did not investigate the memory impact of enabling the cache, which may be relevant in systems with limited memory. This is left for future work.

\section{Analytic time complexity}
\label{sec:analytic}
    \begin{figure}
        \centering
        \includegraphics[width = \columnwidth] {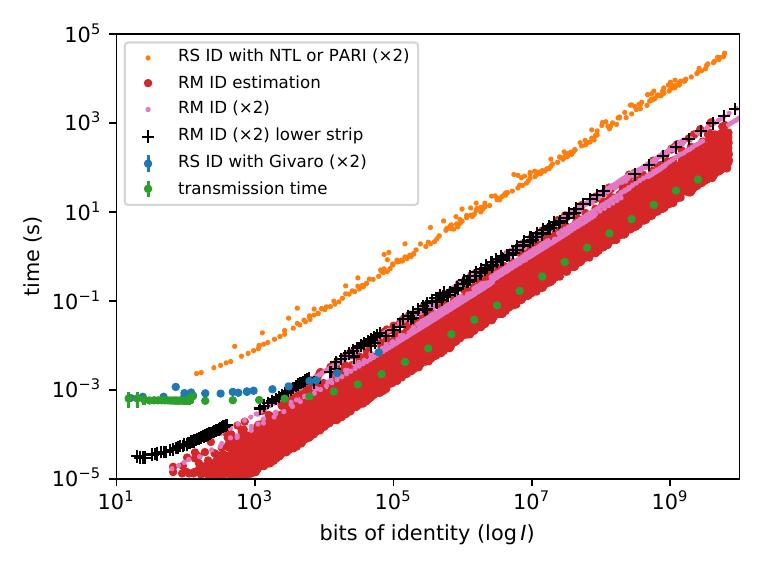}
        \caption{Measured (black and pink) and estimated (red) time cost of Reed-Muller identification, together with the time cost of Reed-Solomon identification (orange and blue) and transmission (green) from~\cite{FTBDLMV21} for comparison.}
        \label{fig:analytic}
    \end{figure}
    We tried to estimate the time spent by the Reed-Muller identification encoder with the goal of estimating semi-analytically the best parameters in terms of time, size and error. 
    However, the analysis of this estimation did not accurately predict the best measured parameter.
    This is explained in detail below together with possible further improvements.
    
    The time estimation was done by simply counting the number of additions and multiplications performed by the recursive implementation of the polynomial.
    Let $t_+(q)$ and $t_*(q)$ be the times of performing one addition or multiplication respectively, and let us assume that exponentiation has the same cost as multiplication.
    The estimated time has a simple recursive relation given by
    \begin{align*}
        C(q,m,0) &= 0
        \qquad
        C(q,1,k) = k t_+(q) + 2k t_*(q)
        \\
        C(q,m,k) &= k t_+(q) +  k t_*(q) + \sum\nolimits_{k'\in[k]} C(m-1,k-k')
        \\
               &= k t_+(q) +  k t_*(q) + \sum\nolimits_{k'\in[k]} C(m-1,k')
    \end{align*}
    When operation time is constant $t_+(q) = t_*(q) = t$ as for $q<2^{16}$, the above cost function satisfies 
    \[C(q,m,k) = t \cdot C(m,k)\]
    where $C(m,k)$ is $C(q,m,k)$ calculated with $t_+ = t_* = 1$.
    This suggests to use the largest field size below $2^{16}$ in order to reduce the error and increase the size of the Reed-Muller identification code, since no penalty is incurred in choosing these fields.
    The analysis can then focus on finding the best parameters $m$ and $k$ that optimize the estimated encoding time $C(m,k)$.
    By induction, the highest order term in $C(m,k)$ is $3\frac{k^m}{m!}$, however, since already the exact computation of $C(m,k)$ did not lead to the desired results, we did not investigate further how well $3\frac{k^m}{m!}$ approximates $C(m,k)$.
    
    The estimated time plotted against the size $\log I$ is shown in the red points in \cref{fig:analytic} for $r,m = 1,\dots, 50$.
    The points form a band with the same slope as measured points (black and pink), suggesting that the bottom of the band could lead to optimized parameters. 
    We divided in strips and the lowest was used for the parameters measured in the black points, which however lie among the slowest points of the measured parameters.
    We take this as an indication that $C(m,k)$ is too simple to give accurate predictions.
    More accurate estimates could be achieved by including the cost of recursion and variables assignment, which is left for future work.

\end{document}